\documentclass[acs,twocolumn,amsmath,amssymb,superscriptaddress,reprint,longbibliography]{revtex4-1}
\usepackage{amsfonts,amsmath,amssymb,bm}
\usepackage{graphicx,graphics,float}
\usepackage{bm}
\usepackage{amssymb}
\usepackage{colordvi}
\usepackage{graphicx}
\usepackage{color}
\usepackage[colorlinks=true,linkcolor=blue,citecolor=blue,urlcolor=blue]{hyperref}%
\usepackage{hyperref}
\usepackage{comment}
\usepackage{harpoon}
\usepackage{pifont}
\newcommand{\be}{\begin{equation}}
\newcommand{\ee}{\end{equation}}
\newcommand{\eps}{\varepsilon}

\begin{document}

\title{Tuning Topological Transitions in Twisted Thermophotovoltaic Systems}

\author{Rongqian Wang}
\address{Institute of Theoretical and Applied Physics, School of Physical Science and Technology \& Collaborative Innovation Center of Suzhou Nano Science and Technology, Soochow University, Suzhou 215006, China.}

\author{Jincheng Lu}
\address{Jiangsu Key Laboratory of Micro and Nano Heat Fluid Flow Technology and Energy Application, School of Physical Science and Technology, Suzhou University of Science and Technology, Suzhou, 215009, China}
\address{Institute of Theoretical and Applied Physics, School of Physical Science and Technology \& Collaborative Innovation Center of Suzhou Nano Science and Technology, Soochow University, Suzhou 215006, China.}

\author{Xiaohu Wu}\email{xiaohu.wu@iat.cn}
\address{Shandong Institute of Advanced Technology, Jinan 250100, China}

\author{Jiebin Peng}\email{pengjiebin@hotmail.com}
\address{School of Physics and Optoelectronic Engineering, Guangdong University of Technology, Guangzhou 510006, Guangdong Province, PR China}

\author{Jian-Hua Jiang}\email{jianhuajiang@suda.edu.cn}
\address{Institute of Theoretical and Applied Physics, School of Physical Science and Technology \& Collaborative Innovation Center of Suzhou Nano Science and Technology, Soochow University, Suzhou 215006, China.}

\date{\today}
\begin{abstract}
Twisted bilayer two-dimensional electronic systems give rise to many exotic phenomena and unveil a new frontier for the study of quantum materials. In photonics, twisted two-dimensional systems coupled via near-field interactions offer a platform to study localization and lasing. Here, we propose that twisting can be an unprecedented tool to tune the performance of near-field thermophotovoltaic systems. Remarkably, through twisting-induced photonic topological transitions, we achieve significant tuning of the thermophotovoltaic energy efficiency and power. The underlying mechanism is related to the change of the photonic iso-frequency contours from elliptical to hyperbolic geometries in a setup where the hexagonal-boron-nitride metasurface serves as the heat source and the indium antimonide $p$-$n$ junction serves as the cell. We find a notably high energy efficiency, nearly 53\% of the Carnot efficiency, can be achieved in our thermophotovoltaic system, while the output power can reach to $1.1\times10^4$~W/m$^2$ without requiring a large temperature difference between the source and the cell. Our results indicate the promising future of twisted near-field thermophotovoltaics and paves the way towards tunable, high-performance thermophotovoltaics and infrared detection.
\end{abstract}

\maketitle

As a solid-state renewable energy resource, thermophotovoltaic (TPV) systems have shown great potential in a wide of applications including solar energy utilization and waste heat recovery~\cite{bauer11,chan2013toward,nature22}. The photovoltaic cell, which is separated from a thermal emitter by a vacuum gap in a TPV system, can absorb the thermal radiation from the emitter and converts heat into the electrical energy via the photovoltaic effect~\cite{liao2016efficiently,Tervo2018}. Since the power densities of far-field TPV systems are fundamentally constraint by the Stefan-Boltzmann's law, near-field effects are explored to break the blackbody limit~\cite{RMPBen,molesky2015ideal,st2017hot,papadakis2020broadening}. By reducing the vacuum gap to be comparable to or smaller than the characterized thermal wavelength, the power densities in the near-field TPV (NTPV) systems can far exceed that in the far-field systems through strong near-field coupling.
To effectively match the gap frequency of the photovoltaic cell to the emission spectrum of the emitter, one can exploit the coupling of surface polaritons, e.g., the surface-plasmon polaritons~\cite{tang2020near,zhang2020active}, the surface-phonon polaritons (SPhPs)~\cite{biehs2012hyperbolic,ghashami2018precision}, and the magnetoplasmon polaritons~\cite{Wu2019magneto,PengACS}. Various types of surface polaritons have been extensively studied and demonstrated their ability in enhancing the near-field thermal radiation and improving the performances of the NTPV systems~\cite{wu2012metamaterial,svetovoy2012plasmon, ilic2012overcoming,svetovoy2014graphene,basu2015near,zhao2017high,NC20,mittapally21}. Moreover, it has been shown that the near-field radiative heat transfer can be greatly enhanced by the hybridization effect of polaritons. In particular, the surface-plasmon polaritons in graphene can couple with the SPhPs in hexagonal boron nitride ($h$-BN) films to form hybrid polaritons that assist the photon tunneling~\cite{Bo_PRB,WuIJEM,Small}. Recently, inspired by the pioneering work of Hu {\it et al.},  the heat transfer based on the twisted hyperbolic systems has been explored~\cite{hu2020moire,hu2020topological}.
The emitter/absorber in these systems is made by uniaxial hyperbolic metasurface and can support several types of anisotropic hyperbolic polaritons.
By manipulating the twist angle between the emitter and the absorber, topological transitions can be induced in the dispersion of the polaritons, resulting in the enhancement of the radiative heat transfer~\cite{he2020active,zhou2021polariton}. The existing studies on the near-field thermal management via twist are restricted to the radiative heat transfer, however, a primary concern of topological transitions with the NTPV systems is still missing.

The high-performance of NTPV systems is limited due to the challenges in creating selective thermal emitters with large emittance for photon energies larger than the bandgap energy of the photovoltaic cell and creating photovoltaic cells with perfect absorbing the above-gap thermal radiation~\cite{15PRL,16ACS,20NC}. In our previous works, we designed and optimized graphene-$h$-BN-InSb near-field structures to enhance the performances of the NTPV systems working in the temperature range of common industrial waste heat (400 K to 800 K)~\cite{PRAppliedWang,WangCPL21}. When a thin $h$-BN flake is patterned into a metasurface structure, the two permittivities along the direction perpendicular to the optical axis become anisotropic~\cite{Zhang2007Nano,20NC}. The in-plane hyperbolicity arises in $h$-BN metasurfaces when the real part of two permittivities have different signs, which supports highly confined in-plane hyperbolic phonon polaritons. Such anisotropic polaritons give rise to the enhancement of the photonic local density of states (PLDOS) around the synthetic transverse optical (STO) phonon frequency and enable well near-field matching of polaritons above the gap frequency in InSb~\cite{20NC}.

In this work, we propose to use the $h$-BN metasurface to perform as a selective near-field emitter and narrow bandgap indium antimonide (InSb) $p$-$n$ junction with a $h$-BN metasurface as an effective TPV cell and examine the effects of the twists in both of in-plane and out-of-plane directions on the performances of the NTPV systems. By twisting the optical axis of the $h$-BN metasurface to the $y$-direction, we demonstrate that the out-of-plane topological transitions of the dispersion of SPhPs enables the suppression of the excitation of the SPhPs, which reduces the below-gap radiative heat transfer between the emitter and the cell~\cite{liu2016super,wu18influence,WU21JQSRT}. By manipulating the twist angle between the two $h$-BN metasurfaces, we show that the in-plane topological transitions further affect the output power and energy conversion efficiency of the proposed system.


\begin{figure}
	\begin{center}
		\includegraphics[width=3.5 in]{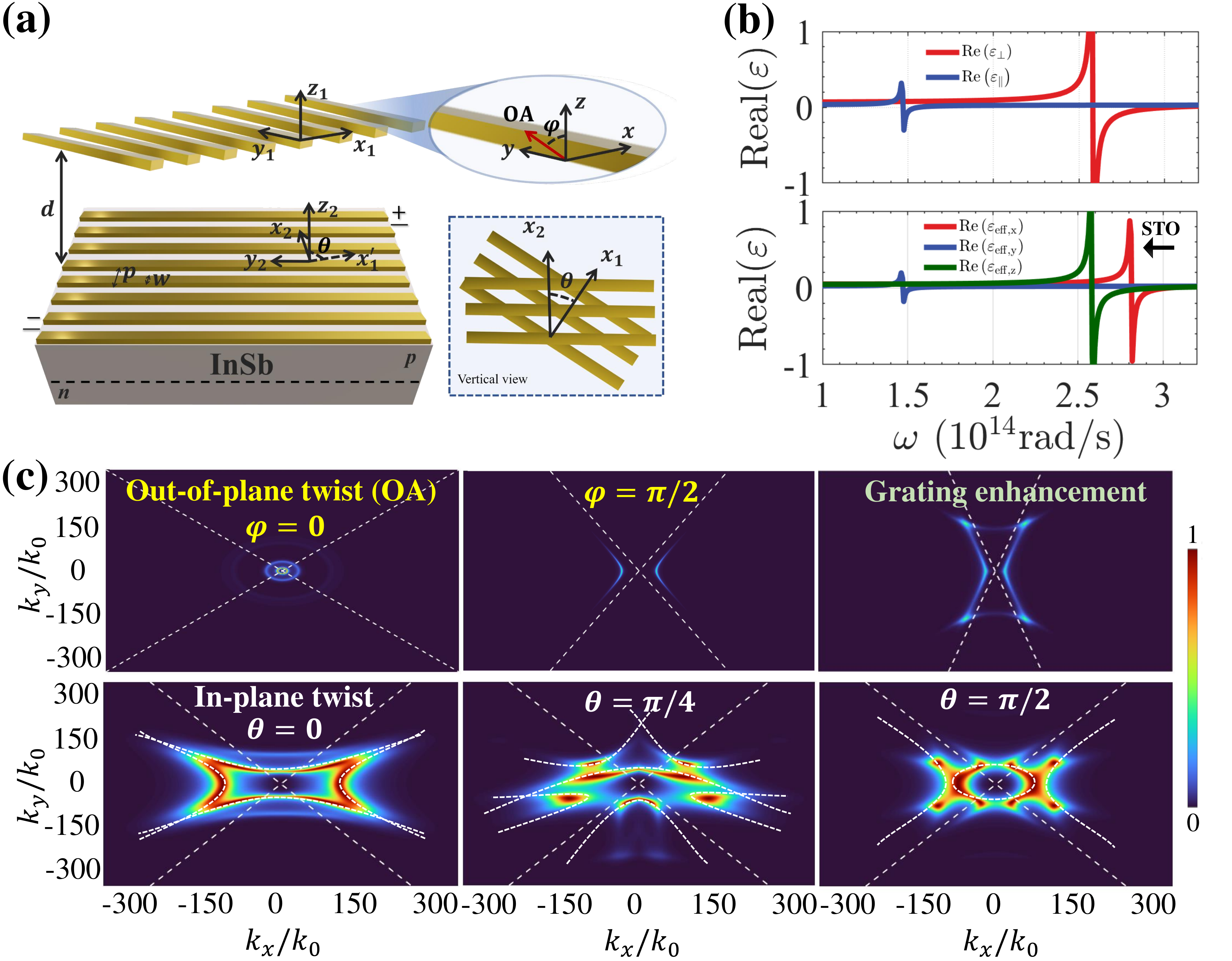}
		\caption{(a) Schematic illustration of the NTPV system. The near-field emitter of temperature $T_{\rm emit}$ is a $h$-BN metasurface based on a grating made of nanoribbons, whose periodicity and stripe width  are $p$ and $w$, respectively. The cell of temperature  $T_{\rm cell}$ is a InSb $p$-$n$ junctiong coated by an identical metasurface of the emitter. The distance between the emitter and the cell is set as $d$. The temperature of the $h$-BN metasurface attached to the InSb cell has the same temperature as the InSb cell. The twisted angle between the main axes of the two gratings is $\theta$. The upper inset: the optic axis of $h$-BN is in the $x$-$z$ plane and is tilted off the $z$-axis by an angle $\varphi$. The right inset: top view of the twisted hyperbolic system at a rotation angle $\theta$ with respect to $x$ axis.
		(b) The upper panel: real parts of the in-plane and out-of-plane permittivities of the $h$-BN; the lower panel: real parts of the permittivities along three principal axes of the $h$-BN metasurface. (c) The topological transitions induced by the out-of-plane (upper panel) and in-plane twists (lower panel).}\label{fig:1}
	\end{center}
\end{figure}
{\it Near-field thermal radiation.--}The proposed NTPV system is illustrated by Fig.~\ref{fig:1}(a), where the near-field emitter with temperature $T_{\rm emit}$ and the cell with temperature $T_{\rm cell}$ are separated by a vacuum gap with distance $d$. The emitter is a $h$-BN metasurface with thickness $t_{\rm BN}$, which consists of a periodic array of $h$-BN nanoribbons. The periodicity and stripe width of the nanoribbons are $p$ and $w$, respectively.
The TPV cell is made of an InSb $p$-$n$ junction, which is coated by an identical $h$-BN metasurface of the emitter. In this NTPV system, in-plane hyperbolic phonon polaritons induced by the in-plane anisotropy of the $h$-BN metasurfaces are excited to enhance the near-field coupling between the emitter and the cell. The effective anisotropic permittivities tensor ${\bf{\eps}}_ {\rm{eff}}$ of the $h$-BN metasurface are described by a modified effective medium model (optic axis in $z$ direction
)~\cite{15PRL,16ACS,20NC}.
\begin{equation}
	\begin{aligned}
		\eps_{\rm{eff},\it{x}}=&\left(\frac{1-\xi}{\eps_{\perp}}+\frac{\xi}{\eps_c}\right)^{-1},\\
		\eps_{\rm{eff},\it{y}}=&\left(1-\xi\right)\eps_{\perp}+\xi\eps_{\rm air},\\
		\eps_{\rm{eff},\it{z}}=&\left(1-\xi\right)\eps_{\parallel}+\xi\eps_{\rm air},	
	\end{aligned}
\end{equation}
where $\xi=\left(p-w\right)/p$ is the filling factor. $\eps_c=\frac{2p}{\pi t_{\rm BN}}\rm{ln}\left[\rm{csc}\left(\frac{\pi}{2}\xi\right)\right]$ is a nonlocal correction parameter taking into account the near-field coupling of adjacent ribbons and the corresponding nonlocality~\cite{15PRL,16ACS,20NC}. $\eps_{\perp}$ and $\eps_{\parallel}$ are the in-plane and out-of-plane permittivities of $h$-BN, respectively~\cite{caldwell14sub,dai14tunable,li18infrared}. $\eps_{\rm{eff}}$ is a diagonal matrix with $\eps_{\rm{eff},\it{x}},\eps_{\rm{eff},\it{y}},\eps_{\rm{eff},\it{z}}$. For convenience, we define two kinds of angles: out-of-plane twist angle as $\varphi$ which denotes the angle between $z$ direction and optic axis  of $h$-BN and describes the rotation of optic axis in $y$-$z$ plane; in-plane twist angle as $\theta$ which denotes the angle between two parallel $h$-BN meta-surface and describes the relative rotation in $x$-$y$ plane.

The exchanged radiative heat flux between the emitter and the cell can be calculated based on the fluctuational electrodynamics ~\cite{polder1971theory,pendry1999radiative}:
\begin{align}
\begin{split}
	&Q_{\rm rad}(T_{\rm emit},T_{\rm cell},\omega,\Delta\mu) \\
	&=\frac{\Theta_{1}(T_{\rm emit},\omega)}{{4\pi^2}}\int_{0}^{2\pi}\int_{0}^{\infty}\zeta(\omega,k,\phi)kdkd\phi \\
	&-\frac{\Theta_{2}(T_{\rm cell},\omega,\Delta \mu )}{4\pi^2}\int_{0}^{2\pi}\int_{0}^{\infty}\zeta(\omega,k,\phi)kdkd\phi, \label{Qrad}
\end{split}
\end{align}
where $\Theta_1\left(T_{\rm emit},\omega\right)={\hbar\omega}/[\exp{\left(\frac{\hbar\omega}{k_{\rm B}T_{\rm emit}}\right)}-1]$ and $\Theta_2\left(T_{\rm cell},\omega,\Delta\mu\right)={\hbar\omega}/{[\exp{\left(\frac{\hbar\omega-\Delta\mu}{k_{\rm B}T_{\rm cell}}\right)}-1]}$ are the Planck mean oscillator energies of blackbody at temperature $T_{\rm emit}$ and $T_{\rm cell}$, respectively. $\Delta\mu$ is the electrochemical potential difference across the $p$-$n$ junction for $\omega>\omega_{\rm gap}$, $\phi$ is the azimuthal angle. $\zeta\left(\omega,k,\phi\right)$ is the photon transmission coefficient that describes the probability of photons to tunnel across the vacuum gap, which can be expressed as~\cite{wu18influence}:

\begin{widetext}
\begin{align}
	\zeta(\omega,k,\phi)=\left\{
	\begin{aligned}
	&{\rm{Tr}\left[\left(I-{ \bf{R}}_{\rm {cell}}^{\dagger}\bf{R}_{\rm {cell}}-\bf{T}_{\rm {cell}}^{\dagger}\bf{T}_{\rm {cell}}\right){\bf{D}}\left(I-{\bf{R}}_{\rm {emit}}^{\dagger}{\bf{R}}_{\rm {emit}}-{\bf{T}}_{\rm {emit}}^{\dagger}{\bf{T}}_{\rm {emit}}\right){\bf{D}}^{\dagger}\right]}, {k<k_0 }\\
	&{\rm{Tr}\left[\left({\bf{R}}_{\rm{cell}}^{\dagger}-{\bf{R}}_{\rm {cell}}\right){\bf{D}}\left({\bf{R}}_{\rm {emit}}-{\bf{R}}_{\rm {emit}}^{\dagger}\right){\bf{D}}^{\dagger}\right]} {e^{-2\left|k_z\right|d}}, k>k_0 \label{photon tunneling}
\end{aligned}
\right.
\end{align}
\end{widetext}
Where, $k_0=\omega/c$ is the wavevector in vacuum with c the speed of light in vacuum. $k=\sqrt{k_x^2+k_y^2}$ and $k_z=\sqrt{k_0^2-k^2}$ are the wavevector components parallel and perpendicular to the $x$-$y$ plane in vacuum, respectively. $k_x$ and $k_y$ are the wavevector components along the $x$ axis and $y$ axis, respectively. Note that $\rm{Tr\left(\cdot\right)}$ takes the trace of a matrix and ``$\dagger$'' denotes conjugate transpose. $\bf{I}$ is a $2\times2$ unit matrix and  $\rm{\bf{D}} =\left(\rm{\bf{I}}-\rm{{\bf{R}}_{emit}}\rm{{\bf{R}}_{cell}}{\it e^{{\rm 2}ik_zd}}\right)^{-1}$ is the Fabry-P{\'e}rot-like denominator matrix describing the multiple scattering between the two interfaces of the emitter and the cell. $\rm \bf{R}_{\it j}$ and $\rm \bf{T}_{\it j}\ \left(\it{j}=\rm{emit,cell}\right)$ are the reflection and transmission coefficient matrices at the interface between the emitter (cell) and the air, respectively. These coefficients can be obtained using a modified $4\times4$ transfer matrix method~\cite{wu18influence,PengACS}.

The permittivities of $h$-BN and the proposed $h$-BN metasurfaces are presented in Fig.~\ref{fig:1}(b). As shown in the upper panel, the in-plane permittivity $\eps_{\perp}$ is negative in the frequency range from $1.5\times10^{14}\, \rm{rad/s}$ to $1.6\times10^{14}\, \rm{rad/s}$; the out-of-plane permittivity $\eps_{\parallel}$ is negative in the range of $2.58\times10^{14}\,\rm{rad/s}- 3.03\times10^{14}\, \rm{rad/s}$ for natural $h$-BN. The negativeness of the permittivities defines two frequency regions with strong light-matter interaction, respectively termed as type I and type II hyperbolic regions. The reststrahlen effects occuring in these two regions lead to strong reflection and suppressed transmission which are crucial for the enhancement of near-field thermal radiation~\cite{PRAppliedWang}.

In the lower panel of Fig.~\ref{fig:1}(b), we observe that a significant difference between the in-plane permittivities $\eps_{\rm{eff},\it{x}}$ and $\eps_{\rm{eff},\it{y}}$ is conspicuously distinguished when the $h$-BN layer is patterned into a metasurface structure. For the $h$-BN being constructed as a metasurface, the STO resonance at $\omega=2.78\times10^{14} \, {\rm rad/s}$ emerges in the middle of the upper reststranlen band, which is originated from the strong collective near-field coupling of the dipolar phonon polaritons resonance of the individual $h$-BN ribbons~\cite{20NC}. The emergence of the STO resonance brings about extra hyperbolic band (with $\eps_{\rm{eff},\it{y}}>0$,$\eps_{\rm{eff},\it{x}}<0$), as indicated by the red solid line of Fig.~\ref{fig:1}(b). Due to the STO, the PLDOS on the metasurface differs dramatically from the natural $h$-BN. As demonstrated in Ref.~[\onlinecite{20NC}], the $h$-BN thin film only exhibits one PLDOS peak within the reststranlen band while the metasurface exhibits two PLDOS peaks located on either side of the STO frequency. Since the high PLDOS indicate the strong anisotropic phonon polaritons, which are efficient carriers in the near-field tunneling, the $h$-BN metasurface may serve as a robust near-field emitter for the NTPV system.

\begin{figure*}
	\centering\includegraphics[width=7 in]{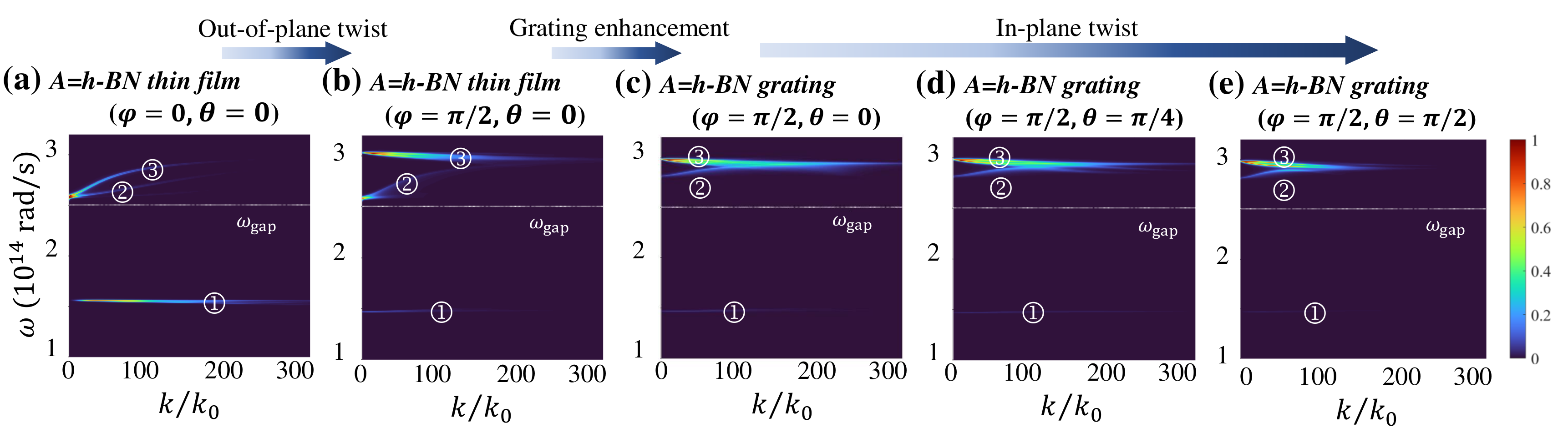}
	\caption{The photon transmission coefficient $\zeta$ as functions of $\omega$ and $k$. (a) and (b) Out-of-plane twist on the photon transmissions of NTPV systems. (c)-(e) In-plane twist on the NTPV system based on $h$-BN gratings. The parameters: the thickness of both $h$-BN thin film and metasurface is $t_{\rm BN}=10$ nm, the periodicity and the filling factor of the $h$-BN metasurface are $p=100$ nm and  $\xi=0.4$, respectively. The temperatures of the emitter and the cell are set at $T_{\rm emit}=450$ K and $T_{\rm cell}=320$ K, respectively. The vacuum gap distance is $d=20$ nm.}\label{fig:2}
\end{figure*}

{\it Twist-induced topological transition.--} To demonstrate the physical mechanisms of twist-induced topological transition in NTPV system, Fig.~\ref{fig:1}(c) examines the photon transmission coefficient $\zeta$ in the $k_x$-$k_y$ plane with fixed frequency $\omega$. The upper and lower panels show the out-of-plane and in-plane twists induced topological transitions, respectively. The bright branches indicate the high photon transmission and the dash-lines are numerical solution of implicit function about the Fabry-P{\'e}rot-like denominator matrix $\bf{D}$, i.e., $\mathrm{Tr}({\bf{D}}^{-1})=0$. With the optical-axis lying in the $z$-direction (upper panel), the energy transmission coefficient exhibits rotational symmetry in the $k_x$-$k_y$ plane arising from the in-plane isotropy of $h$-BN, which indicates the contribution of the elliptical surface phonon polaritons (ESPhPs). The ESPhPs supported by the $h$-BN thin films are excited and contribute to the photon tunneling between the emitter and the cell. After twisting the optical axis to the $y$-direction, the photon transmission coefficient becomes hyperbolic due to the in-plane anisotropy. It is demonstrated that the hyperbolic surface phonon polaritons (HSPhPs) are excited after optical axis twisting and there is out-of-plane twist induced topological transition. In addition, owing to the emergence of the new hyperbolic band through the construction of grating metasurface, more high-$k$ HSPhPs can be excited and thus enhance the photon transmission. The lower panel of Fig.~\ref{fig:1}(c) presents another in-plane twist induced topological transition at resonance frequency: when the twist angle $\theta$ between two metasurfaces changes from zero to $\pi/2$, the bright branch of the photon transmission coefficient change markedly from hyperbolic (open) to elliptical (closed).
We also find that there are band splitting (from 0 to $\pi/4$) and band degeneracy (from $\pi/4$ to $\pi/2$) due to the $\theta$-dependet near-field coupling. Meanwhile, the excitation of the high-$k$ HSPhPs are suppressed, and the low-$k$ ESPhPs dominate the photon tunnleing, resulting in the reduction of radiative heat flux.

To show the tuning effects of the twists-induced topological transitions in NTPV system, Fig.~\ref{fig:2} investigates the photon transmission coefficients in $\omega-k$ plane. For convenience, we select three significant modes in the enhanced transmission for these systems, respectively denoted by modes \ding{172},\ding{173} and \ding{174}. Figure~\ref{fig:2}(a) shows that the photon transmission is enhanced in the two hyperbolic regions of $h$-BN. The enhanced transmission below the gap frequency of InSb $p$-$n$ junction (mode \ding{172}, about $1.55\times10^{14}\, \rm rad/s$) is due to the strong coupling of the ESPhPs supported by the $h$-BN thin films of the emitter and the cell. The high transmission above the gap frequency of InSb $p$-$n$ junction (modes \ding{173}, about $2.62\times10^{14}\ \rm rad/s$ and \ding{174}, about $2.81\times10^{14}\ \rm rad/s$) is originated from the resonant coupling between the ESPhPs in $h$-BN and the electron-hole excitations in the InSb $p$-$n$ junction.

As shown in Fig.~\ref{fig:2}(b), there are three salient features of topological transition caused by the twist of the optical axis: (\romannumeral1) the below-gap ESPhPs (mode \ding{172}, $\approx1.55\times10^{14}\ \rm rad/s$) is significantly weakened; (\romannumeral2) one above-gap transmission mode \ding{173}($\approx 2.73\times10^{14}\ \rm rad/s$) can be transformed from ESPhPs to HSPhPs with reduced transmission probability and wave vector range $k_{\rm max}<100k_0$; (\romannumeral3) another ESPhPs (mode \ding{174}$\approx 2.99\times10^{14}\ \rm rad/s$) has emerged due to the in-plane anisotropy. Due to the strong coupling of the emerged ESPhPs between the emitter and the cell, the photon transmission around this mode is much enhanced compared to the case in Fig.~\ref{fig:2}(a), with increased transmission probability and wave vector range $k_{\rm max}\approx180k_0$. These results are consistent with Ref.~\cite{wu18influence}.


The near-field coupling between HSPhPs and emerged ESPhPs can be enhanced by the grating effects.  For mode \ding{173}(about $2.83\times10^{14}\ \rm rad/s$), the in-plane hyperbolic dispersion ( $\eps_x=-47.9263+14.5097i$, $\eps_y=0.0003+0.1102i$) are formed by near-field coupling of the phonon polaritons in individual $h$-BN nano-ribbons~\cite{20NC}.
For mode \ding{174} (about $2.96\times10^{14}\, \rm rad/s$), there is a elliptical in-plane dispersion with $\eps_x=-2.1815+0.2079i$ and $\eps_y= -2.0932+0.0002i$ for the phonon polaritons.
As presented in Fig.~\ref{fig:2}(c), when the $h$-BN thin film is patterned into a metasurface structure (with $p=100$ nm, $\xi=0.4$), the above-gap HSPhPs \ding{173}($\approx 2.83\times10^{14}\, \rm rad/s$) and emerged ESPhPs \ding{174}( $\approx2.96\times10^{14}\, \rm rad/s$) of the NTPV system are further hybridized and form a strong resonant mode at high-$k$ region. From the point of topological transition, we can find that the resonant mode can form a flat band at $k_x-k_y$ plane, i.e. transition modes between ESPhPs and HSPhPs, and the PLDOS can be significantly enhanced at resonant (flat band) condition. Originated from the strong collective near-field coupling (negative $\varepsilon_x$) of individual nanoresonators (Fabry-P{\'e}rot polariton resonances at nanoribbons), the ESPhPs and HSPhPs are both excited and contributes to the enhanced transmission.

As demonstrated in Figs.~\ref{fig:2}(c)-\ref{fig:2}(e), we move to the dependence of in-plane twist on photon transmission $\zeta$, including both the below gap ($\omega<\omega_{\rm gap}$) and above gap ($\omega>\omega_{\rm gap}$). Due to suppression of the HSPhPs induced by the in-plane topological transition, the photon transmission at the coupling mode between modes \ding{173} and \ding{174} becoming weakened and shrink to a narrower wavevector region when the twist angle $\theta$ changes from 0 to $\pi/2$. While for mode \ding{172}, the transmission is slightly enhanced when $\theta=\pi/4$ but suppressed for the twist angle increasing to $\pi/2$. From above spectral analysis, we numerically show an enhancement of photon transmission coefficients through the adjustment of in-plane and out-of-plane twisting. The effects of the near-field couplings on the NTPV systems are elaborated in the following.

\begin{figure*}
	\centering\includegraphics[width=7 in]{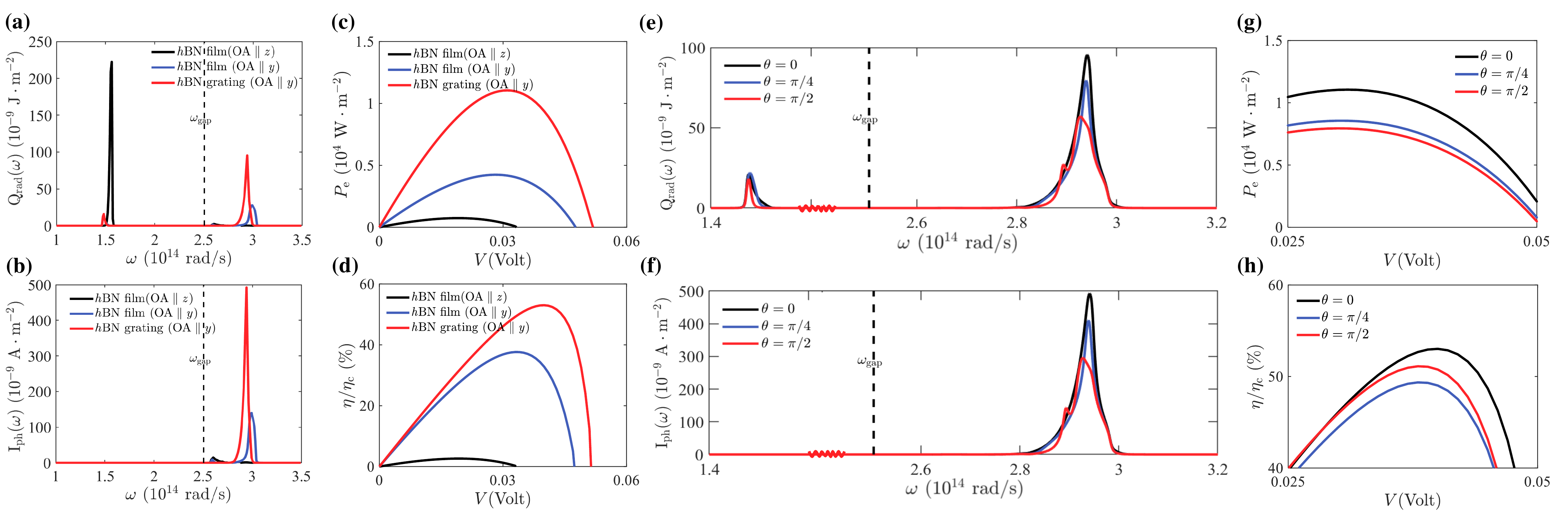}
	\caption{(a)-(d) Out-of-plane twist on the performance of NTPV systems. (a) Radiative heat spectra $Q_{\rm rad}\left(\omega\right)$, and (b) photoinduced current spectra $I_{\rm ph}\left(\omega\right)$ for the $A$-$A$-InSb NTPV systems as a function of $\omega$, where the voltage bias $V$ is at maximum output power for each set-up. (c) Electrical power density $P_e$, and (d) energy$-$conversion efficiency $\eta/\eta_{\rm c}$ in unit of Carnot efficiency ($\eta_{\rm c}$) of the $A$-$A$-InSb NTPV systems as a function of voltage bias $V$.
	(e)-(h) In-plane twist on the performance of NTPV systems.
	(e) $Q_{\rm rad}\left(\omega\right)$, and (f) $I_{\rm ph}\left(\omega\right)$ for the NTPV system based on $h$-BN gratings as a function of $\omega$ at different twisting angles $\theta$, where $V=0.03$ Voit. (g) $P_e$, and (h)  $\eta/\eta_{\rm c}$ for the NTPV system based on $h$-BN gratings as a function of voltage bias $V$ at different twisting angles $\theta$. The parameters: $T_{\rm emit}=450$ K, $T_{\rm cell}=320$ K, $d=20$ nm, $t_{\rm BN}=10$ nm, $p=100$ nm, and $\xi=0.4$.}\label{fig:3}
\end{figure*}


{\it The topological transitions on NTPV systems.--}
Based on the detailed balance analysis, the actual electric current density $I_{\rm e}$ of the TPV cell is derived as ~\cite{shockley1961detailed}
\begin{equation}
\begin{aligned}
I_{\rm e}=(I_{\rm ph}-I_0[\exp(eV/k_{\rm B}T_{\rm cell})-1])\eta_{\rm QE}\equiv {I_{\rm e,max}}{\eta_{\rm QE}},  \label{Ie}
\end{aligned}
\end{equation}
where $V=\Delta\mu/e$ is the operating voltage of the cell~\cite{shockley1961detailed}, $I_{\rm e,max}$ is the maximum electric current density and $\eta_{\rm QE}$ is the mean quantum efficiency~\cite{bauer11}. Here, we consider the ideal situation, i.e., $\eta_{\rm QE}=100\%$. $I_0$ is the reverse saturation current density~\cite{lim2015graphene}, whereas $I_{\rm ph}$ is photo-generation current density, which arises from the radiation absorption of the cell in the frequency range above the band gap
\begin{equation}
	\begin{aligned}
		& I_{\rm ph}=e\int_{\omega_{\rm gap}}^{\infty}\frac{Q_{\rm rad}\left(T_{\rm emit},T_{\rm cell},\omega,\Delta\mu\right)}{\hbar\omega}d\omega, \label{Iph}
	\end{aligned}
\end{equation}

The output electric power density $P_{\rm e}$ of the NTPV system is defined as the product of the net electric current density and the voltage bias,
\begin{equation}
\begin{aligned}
P_{\rm e}= -I_{\rm e}V, \label{epower}
\end{aligned}
\end{equation}
and the incident radiative heat flux is given by the integral of the radiative heat flux defined in Eq.~\eqref{Qrad}
\begin{equation}
\begin{aligned}
&Q_{\rm inc}=\int_{0}^{\infty}Q_{\rm rad}\left(T_{\rm emit},T_{\rm cell},\omega,\Delta\mu\right)d\omega \\
&=\int_{0}^{\infty}\frac{d\omega}{\left(2\pi\right)^3}\Theta_{1}\left(T_{\rm emit},\omega \right)\int_{0}^{2\pi}\int_{0}^{\infty}\zeta\left(\omega,k,\phi\right)kdkd\phi  \\
&-\int_{\omega_{\rm gap}}^{\infty}\frac{d\omega}{\left(2\pi\right)^3}\Theta_{2}\left(T_{\rm cell},\omega,\Delta\mu\right)\int_{0}^{2\pi}\int_{0}^{\infty}\zeta\left(\omega,k,\phi\right)kdkd\phi  \\
&-\int_{0}^{\omega_{\rm gap}}\frac{d\omega}{\left(2\pi\right)^3}\Theta_{1}\left(T_{\rm cell},\omega\right)\int_{0}^{2\pi}\int_{0}^{\infty}\zeta\left(\omega,k,\phi\right)kdkd\phi. \label{Qinc}
\end{aligned}
\end{equation}
The integrand functions of frequency in Eq.~\eqref{Iph} and Eq.~\eqref{Qinc} are defined as the spectral function $Q_{\rm rad}(\omega)$ and $I_{ph}(\omega)$, respectively. The energy efficiency $\eta$ is given by the ratio between the output electrical power density $P_{\rm e}$ and incident radiative heat flux $Q_{\rm inc}$~\cite{Zhang2007Nano},
\begin{equation}
\begin{aligned}
\eta= \frac{P_{\rm e}}{Q_{\rm inc}}.\label{effi}
\end{aligned}
\end{equation}

Figures~\ref{fig:3}(a)-\ref{fig:3}(b) reveal the spectral functions $Q_{\rm rad}(\omega)$ and $I_{\rm{ph}}(\omega)$ of the three different NTPV systems ,i.e., $h$-BN film with optical axis being parallel to $z$-axis (black line), $h$-BN film with twisted optical axis being parallel to $y$-axis (blue line), and $h$-BN grating with twisted optical axis being parallel to $y$-axis (red line). All three lines in Fig.~\ref{fig:3}(a) shows that the radiative heat spectra $Q_{\rm rad}(\omega)$ is strengthened in the two hyperbolic bands of $h$-BN for all three NTPV systems. But the black line shows that the incident heat spectra below the gap frequency of InSb is much higher than the above one for $h$-BN thin film without twisted optical axis. When the optical axis is twisted $\pi/2$ to the $y$-direction (the two $A$-$A$-InSb NTPV systems with $A$=$h$-BN thin film and $A$=$h$-BN grating), this high heat spectra decreases dramatically due to the suppression of ESPhPs caused by the twist of the optical axis. The similar phenomena has been studied in Ref.~\cite{wu18influence}.
For the system with $h$-BN thin film (with the optical axis lies in the $y$-direction), the incident heat spectra above the gap frequency $\omega_{\rm gap}$ is enhanced predominantly comparing with the blow one attributing the success to the topological transition of ESPhPs to HSPhPs and the emerged ESPhPs modes. For the NTPV system with $A$=$h$-BN metasurface, due to the strong coupling between ESPhPs and HSPhPs, a higher incident heat spectra exhibits above the gap frequency of InSb. Consistent with the incident heat spectra, the photon-induced current of the system with $A$=$h$-BN metasurface is much larger than the both two systems with $h$-BN thin films.

Indeed, from Eq.~\eqref{Iph}, it is clear to find {\it only} the contribution of above the band gap, i.e., $\omega>\omega_{\rm gap}$, can improve the performances of NTPV cells~\cite{messina2013graphene}. In contrast, the enhancement below the band gap leads to reduced energy efficiency, since these photons are useless for energy conversion.
The high transmission region above the band gap is mainly due to the resonant coupling between the ESPhPs(HSPhPs) and the electron-hole excitation in the InSb junction. The system with $A$=$h$-BN grating possesses the significantly increased above-gap incident heat spectra and the dramatically decreased below-gap spectra, shown in Figs.~\ref{fig:3}(a) and \ref{fig:3}(b).
This finally brings about the excellent performance of the system with $A$=$h$-BN grating, with the highest output power about $1.1\times10^4\ \rm W/m^2$ and the energy efficiency nearly 53\% of the Carnot efficiency [see Figs.~\ref{fig:3}(c) and ~\ref{fig:3}(d)]. On the contrary, the highest incident heat spectra below the band gap and the lowest heat spectra above the band gap leads to the lowest energy efficiency and output power of the system $A$=$h$-BN thin film (with the optical axis lies in the $z$-direction)[see the dark solid line in Figs.~\ref{fig:3}(c) and ~\ref{fig:3}(d)]. The moderate performance in the system $A$=$h$-BN thin film (with the optical axis lies in the $y$-direction) is originated from the enhanced above-gap incident heat spectra and the decreased below-gap spectra compared to that of the system $A$=$h$-BN thin film (with the optical axis lies in the $z$-direction).

Eventually, we turn to investigate the effects of the in-plane twist on the performances of the the NTPV system based on $h$-BN gratings [Figs.~\ref{fig:3}(e)-\ref{fig:3}(h)]. It is evidently to observe that the radiative heat spectra $Q_{\rm rad}$ above the gap decreases with the enlargement of twist angle $\theta$ [in Fig.~\ref{fig:3}(e)], so that it leads to the reduction of the photoinduced current spectral functions using Eq.~\eqref{Iph}, see Fig.~\ref{fig:3}(f), which is consistent with the previous analysis.

At this point, it becomes transparent that the optimal electric output power and efficiency of the NTPV system without in-plane twist are much better than those with twist. The NTPV system without twist benefits from the narrowband nature of photon absorption, which gives rise to improved energy efficiency as compared with broadband photon absorption, as proved in Ref.~\cite{Jiang2018Near}. Here we have to point that for the below-gap heat spectra, the energy efficiency exhibits a nonmonotoinc dependency versus the twisting angle $\theta$. According to Eq.~\eqref{effi}, the reduced power divided by the increased incident radiation ultimately leads to reduced efficiency (about 49\% of the Carnot efficiency) for $\theta=\pi/4$. Likewise, the reduction of output power and input radiation contemporaneously produces a moderate efficiency for $\theta=\pi/2$.




{\it {Summary and outlook}.}- In this work, we propose a NTPV system based on $A$=$h$-BN gratings where twisting trigger dramatic changes in the optical properties. In particular, the below-absorption-gap (wasted) radiative heat transfer can be suppressed by twisting the optical axis of $h$-BN into the $y$-axis. As a consequence, the performance of the NTPV system is significantly improved. The energy efficiency can reach to nearly $53\%\eta_C$ while the heat source and the photovoltaic cell are kept at a relatively low temperature difference ($130\, \rm K$). The underlying mechanism is uncovered as the topological transitions in the near-field coupled photonic systems, specifically, in the forms of SPhPs. Meanwhile, the output power can also be tuned by twisting the $h$-BN metasurface. Our study paves the pathway toward high-performance TPV systems based on twisted photonics.

{\it {Acknowledgements.}}-We thank the support from the funding for Distinguished Young Scientist from the National Natural Science Foundation of China (Grant Nos. 12125504, 12074281, 12074281, 12047541, 12074279, and 52106099), the Major Program of Natural Science Research of Jiangsu Higher Education Institutions (Grant No. 18KJA140003), the Jiangsu specially appointed professor funding, and the Academic Program Development of Jiangsu Higher Education (PAPD), the China Postdoctoral Science Foundation (Grant No. 2020M681376), the faculty start-up funding of Suzhou University of Science and Technology, and Jiangsu Key Disciplines of the Fourteenth Five-Year Plan (Grant No. 2021135).

\bibliography{ref_hgInSb}

\end{document}